\documentclass[aps,pra,epsfig]{revtex4}

\usepackage[dvips]{graphics}
\usepackage{graphicx}
\usepackage{amsfonts}
\usepackage{amssymb}
\usepackage{amsmath}

\begin{document}

\title{Effect of Random On-Site Energies on the Critical Temperature of a 
Lattice Bose Gas}
\author{Luca Dell'Anna$^1$, Stefano Fantoni$^{1,2,3}$, Pasquale Sodano$^{4}$, and Andrea Trombettoni$^{1,2}$} 
\affiliation{$^1$ International School for Advanced Studies, 
Via Beirut 2/4, I-34014, Trieste, Italy\\
$^2$ INFN, Sezione di Trieste\\
$^3$ INFM, CNR-DEMOCRITOS National Supercomputing Center, Trieste, Italy\\
$^4$ Dipartimento di Fisica and INFN, Sezione di Perugia,  
Universit\`a di Perugia, Via A. Pascoli, 
I-06123, Perugia, Italy}

\begin{abstract}
We study the effect of random on-site energies on the critical temperature of a non-interacting 
Bose gas on a lattice. In our derivation the on-site energies are distributed according a Gaussian 
probability distribution function having vanishing average and variance $v_o^2$. 
By using the replicated action obtained by averaging on the disorder, 
we perform a perturbative expansion for the Green functions of the disordered system. 
We evaluate the shift of the chemical potential induced by the 
disorder and we compute, for $v_o^2 \ll 1$, the critical temperature for condensation. 
We find that, for large filling, disorder slightly enhances the critical temperature for condensation. 
\end{abstract}
\maketitle

\section{Introduction}

The study of the effect of disordered potentials on quantum particles is, since decades, an important 
field of research, 
one of the main motivations being the understanding of how electrons in disordered systems 
localize \cite{anderson58,ramakrishnan86}. The investigation of disorder effects on bosonic systems, 
motivated by experiments of adsorption of $^4$He in porous media \cite{kiewiet75,hertz79}, has been 
also very active, especially in relation to the issue of understanding how 
superfluidity properties are modified in random environments 
\cite{fisher89,huang92,giorgini94}. In this respect, ultracold bosonic trapped gases 
\cite{pethick02,stringari03} provide a good experimental setup to study the effects of 
disorder on bosonic systems. In an ultracold Bose gas, 
disorder can be induced both by a laser speckle
\cite{lye05,clement05} or by an incommensurate bichromatic potential \cite{fallani07} (i.e.,
an auxiliary, incommensurate, lattice added to an optical lattice). These techniques recently allowed 
for the experimental observation of Anderson localization for matter waves 
in a random potential \cite{roati08,billy08}. 

An important resource to control the properties of ultracold bosons is 
provided by the possibility of superimposing on them optical lattices \cite{morsch06}, allowing for a 
fine tuning of the ratio between kinetic and interaction energies. The effective  
Hamiltonian for bosons in a deep optical lattice is the Bose-Hubbard Hamiltonian 
\cite{fisher89,jaksch98}, 
i.e., a tight-binding model characterized by a kinetic term describing the hopping of bosons 
between neighbouring sites of the lattice and an interaction term proportional to the $s$-wave 
scattering length between bosons. In absence of disorder, the Bose-Hubbard Hamiltonian has 
a superfluid (Mott insulator) ground-state for large (small) ratio 
between kinetic and interaction energies. 
Random on-site disorder together with a strong interparticle interaction between lattice bosons 
is expected to induce a Bose glass phase \cite{giamarchi88,fisher89} which has been recently studied 
both from a theoretical  
\cite{gimperlein05,rey06,krutitsky06,roscilde07,sengupta07,buonsante07,roscilde08,deng08,roux08,morrison08,krutitsky08,bissbort08}
and experimental \cite{fallani07} side. In the opposite limit of very weak interaction, at $T=0$ 
the Anderson localization has been studied \cite{roati08}: the study of this limit is feasible 
thanks to the possibility of tuning the $s$-wave scattering length $a$ of a $^{39}$Rb Bose gas 
with high precision, and setting it almost to zero \cite{roati07}. 
When an optical lattice is superimposed to an ideal Bose gas, 
for large values of the laser power,
the system is described by a boson-hopping model, i.e., the Bose-Hubbard model 
\cite{fisher89,jaksch98} without the interaction term. 

In \cite{dellanna08} the thermodynamic properties of a Bose gas in 
a disordered lattice have been analyzed and the shift 
of the critical temperature for condensation induced by the disorder was determined. 
For a three-dimensional lattice it has been found that the shift of the 
critical temperature depends on the filling, i.e., on the number of particles per lattice site. 
When disorder affects the hopping rates between neighbouring sites, 
the critical temperature $T_c$ is enhanced for large 
filling $f$ and the shift does not sensibly 
depend on $f$ \cite{dellanna08}; at variance, for small $f$, $T_c$ decreases. In presence of random on-site 
energies, $T_c$ increases for large filling, but much less than for for bond-disordered lattices, 
resulting in very small shift 
of $T_c$ for small disorder. These results should be compared with the findings for a continuous 
(i.e., without optical lattice) Bose gas in presence of a disordering potential 
\cite{lopatin02,zobay06,timmer06,falco07,yukalov07}: without any confining potential, it has been 
shown that the critical temperature decreases with disorder \cite{lopatin02}. 
In Refs.~\cite{yukalov07,yukalov07-a} the properties, 
both at $T=0$ and at finite temperature, of a continuous 
Bose gas in a random potential were investigated in the frame of a self-consistent 
mean-field approach. The shift $\delta T_c$ of the critical temperature 
has been computed in a Hartree-Fock approach by Lopatin and Vinokur \cite{lopatin02} as 
\begin{equation}
\frac{\delta T_c}{T_c^{(0)}}=-{\cal K} \, \frac{ m^3 k_B T_c^{(0)}}{6 \pi^2 \hbar^6 \rho}
\label{continuous}
\end{equation}
where $\rho$ is the density, $m$ the mass of the bosonic atoms and 
$T_c^{(0)}$ the critical temperature for condensation in absence of disorder. 
In Eq.~(\ref{continuous}) ${\cal K}$ is the strength of the disorder: the disorder potential 
$u(\vec{r})$ has averages $\overline{u(\vec{r})}=0$ and 
$\overline{u(\vec{r},\vec{r}^\prime)}={\cal K} \delta(\vec{r}-\vec{r}^\prime)$ 
(the bar denotes the average all disorder configurations). The result (\ref{continuous}) has been 
confirmed in \cite{zobay06} by 
one-loop Wilson renormalization-group calculations. 
The shift of the critical temperature for the continuous Bose gas has been also recently computed 
in \cite{falco07} using the Popov method, 
obtaining a value for the relative shift $\delta T_c / T_c^{(0)}$ which differs 
by a factor of $1/2$ from the result (\ref{continuous}). A discussion of the reasons for 
such difference is contained in \cite{falco07}. 
In presence of an optical lattice, the decrease of the critical temperature is found for small filling, 
for which the result (\ref{continuous}) is retrieved \cite{dellanna08}: however, 
on the lattice, different behaviours are possible as a result of the interplay between discreteness
and disorder \cite{dellanna08}. 

The aim of this paper is to provide details of the results presented in \cite{dellanna08}, 
focusing on the physically relevant case of random on-site energies. We shall present a detailed derivation 
of the computation of the Green functions at the first order in the disorder parameter, and 
discuss the contributions from higher order terms. 
The plan of the paper is the following: in Section II 
the model Hamiltonian is presented and the results in absence of disorder are 
reviewed. In Section III 
the computation of the replicated action is presented. In Section IV, the shift of the chemical 
potential induced by random on-site energies is computed by a perturbative expansion of the Green 
functions and the resulting shift of the critical temperature is determined and discussed. 
Our conclusions are in Section V.

\section{The Model Hamiltonian}

When there are random on-site energies, 
non-interacting bosons on a three-dimensional lattice are described by the Hamiltonian
\begin{equation}
\hat{H}=-t \sum_{\langle i,j \rangle} \left( 
\hat{b}^{\dag}_i \hat{b}_j + \hat{b}^{\dag}_j \hat{b}_i \right) +
\sum_i \epsilon_i \hat{b}^{\dag}_i \hat{b}_i.
\label{HAM}
\end{equation}
In Eq.~(\ref{HAM}) $t$ is the tunneling rate between neighboring sites, 
the lattice sites are denoted by $i,j$, 
the operator $\hat{b}_i$ ($\hat{b}_i^\dag$) destroys (creates) a boson in the lattice site 
$i$ and the sum is on all the distinct pairs of neighbouring sites. 
The number of sites is denoted by $N_S$: for a cubic lattice with of linear 
size $L$, $N_S=L^3$. 
The total number of particles is $N_T$ and 
the filling (i.e. the average number of particles per site) is
given by
\begin{equation}
f=\frac{N_T}{N_S}.
\label{filling}
\end{equation}

In the Hamiltonian (\ref{HAM}), the random on-site energies are accounted for by the $\epsilon_i$'s which 
are regarded as random variables with vanishing average and 
variance $v_o^2 t^2$. We shall consider a small disorder ($v_o^2 \lesssim 1$) and we assume that, at each site,  
the disorder has probability distribution $P(\epsilon_i)$ given by
\begin{equation}
P(\epsilon_i)=\frac{1}{\sqrt{2 \pi v_o^2}} \, e^{-\epsilon_i^2/2 v_o^2}.
\label{prob}
\end{equation}

When there is no disorder ($\epsilon_i=0$), the Hamiltonian (\ref{HAM}) reads 
\begin{equation}
\hat{H^{(0)}}=-t \sum_{\langle i,j \rangle} \left( 
\hat{b}^{\dag}_i \hat{b}_j + \hat{b}^{\dag}_j \hat{b}_i \right)
\label{HAM-ord}
\end{equation}
(hereafter, the superscript $^{(0)}$ is used to describe bosons when on-site disorder is absent). The 
critical temperature for condensation $T_c^{(0)}$ can be determined in the usual way 
\cite{pethick02,stringari03}: one introduces the chemical potential $\mu$ to 
enforce the conservation of the total number of particles and 
replaces $\hat{H}^{(0)}$ with $\hat{K}^{(0)}=\hat{H}^{(0)}-\mu \hat{N}$, 
where $\hat{N}=\sum_i \hat{b}^{\dag}_i \hat{b}_i$ 
is the total number operator. Requiring $\langle \hat{N} \rangle=N_T$ and performing 
the thermodynamical limit ($N_S,N_T \to \infty$ at fixed filling $f$), one gets 
\begin{equation}
\int_{BZ} \frac{d{\bf k}}{(2 \pi)^3}\frac{1}{e^{\beta(E_{{\bf k}}-\mu)}-1}= f,
\end{equation} 
where the integral is on the first Brillouin zone and 
\begin{equation}
E_{{\bf k}}=-2t \left( \cos{k_x} + \cos{k_y} + \cos{k_z} \right)
\label{autovalori}
\end{equation}
are the single particle energies (periodic boundary conditions have been assumed). 
The critical temperature $T_c^{(0)}$ 
is then determined by requiring $\mu(T_c^{(0)}) \equiv \mu_c =-6t$, which yields
\begin{equation}
\int_{BZ} \frac{d{\bf k}}{(2 \pi)^3}\frac{1}{e^{\beta_c^{(0)}
(E_{{\bf k}}-\mu_c)}-1}= f
\label{condizione-Tc}
\end{equation}
(where $\beta_c^{(0)}=1/k_B T_c^{(0)}$).

In \cite{dellanna08} it is shown that a reasonable estimate of $T_c^{(0)}$, valid for large filling, 
may be obtained by keeping 
only the lowest order of the Taylor expansion of the exponential in Eq.~(\ref{condizione-Tc}). 
One gets
\begin{equation}
k_B T_c^{(0)} \simeq \frac{6tf}{{\sf W}(1)},
\label{Tc0}
\end{equation}
where ${\sf W}(1) \simeq 1.516386$ is the value 
of the three-dimensional generalized Watson's integral 
\begin{equation}
{\sf W}(z)\equiv 
\int_{BZ} \frac{d{\bf k}}{(2\pi)^3}\,\frac{1}{1-\frac{1}{3z}\sum_{\ell=1}^3\cos k_\ell} 
\label{watson}
\end{equation}
computed in $z=1$ \cite{mattis85}. In \cite{dellanna08} the value of $T_c^{(0)}$ given 
by Eq.~(\ref{Tc0}) is compared with the numerical solution of Eq.~(\ref{condizione-Tc}), showing 
that this estimate is quite good also for intermediate values of the filling: e.g., for $f \gtrsim5$, 
the relative error is less than $5 \%$.

\section{Effective replicated action}

For the Hamiltonian (\ref{HAM}) with random on-site energies,
the partition function can be written as a path integral \cite{schulman}
\begin{equation}
Z=\int \prod_i {\cal D}\varphi_i\left( \tau \right) {\cal D}\varphi_i^\ast\left( \tau \right) e^{-S}
\label{Z-app}
\end{equation}
($\hbar=k_B=1$), where the action $S$ is given by
\begin{equation}
S=\int_0^{\beta} d \tau \left\{ \sum_i \varphi_i^\ast 
\left( \frac{\partial}{\partial \tau} -\mu \right) \varphi_i +
\sum_i \epsilon_i \varphi_i^\ast \varphi_i -
t \sum_{\langle i,j \rangle} \left( 
\varphi^{\ast}_i \varphi_j + \varphi^{\ast}_j \varphi_i \right)
\right\}.
\label{action-app}
\end{equation}
Of course, the partition function (\ref{Z-app}) depends on the on-site energies distribution, i.e., $Z=Z(\{\epsilon\})$. 
The fields $\varphi$'s describe Bose particles.

As it is usual when one deals with a disordered system, one introduces $N$ replicas of the system and, after averaging the action on the 
disorder, performs the limit $N \to 0$ \cite{dedominicis06}. This allows to write down a perturbative expansion of the Green functions. 
Labeling the $N$ replicas by $\alpha=1,\cdots,N$, one has 
\begin{equation}
Z^N(\{\epsilon\})=\int \prod_{i,\alpha} {\cal D}\varphi_i^\alpha\left( \tau \right) 
{\cal D}\varphi_i^{\alpha \ast}\left( \tau \right) 
\exp{\left\{ - \sum_\alpha \int_0^{\beta} d \tau \left[ \sum_i \varphi_i^{\alpha \ast} 
\left( \frac{\partial}{\partial \tau} -\mu +\epsilon_i \right) \varphi_i^\alpha - 
t \sum_{\langle i,j \rangle} 
\left(\varphi^{\alpha \ast}_i \varphi_j^\alpha +\varphi^{\alpha \ast}_j \varphi_i^\alpha \right)
\right] \right\}}.
\label{Z-app-rep}
\end{equation}

The averaged effective partition function is given by 
\begin{equation}
\overline{Z^N(\{\epsilon\})}=\int \prod_{i} d\epsilon_i P(\epsilon_i) Z^N(\{\epsilon\}),
\label{Z-app-rep-eps}
\end{equation}
where $P(\epsilon_i)$ is the Gaussian probability distribution (\ref{prob}). 
After integrating over the on-site energies $\epsilon_i$'s, one gets 
\begin{equation}
\overline{Z^N(\{c\})}=\int \prod_{i,\alpha} {\cal D}\varphi_i^\alpha\left( \tau \right) 
{\cal D}\varphi_i^{\alpha \ast}\left( \tau \right) 
e^{-S_{eff}},
\label{averaged}
\end{equation}
where the effective replicated action $S_{eff}$ is given by 
\begin{eqnarray}
\label{effective-action}
S_{eff}&=&
\sum_{\alpha} \int_{0}^{\beta} d \tau 
\left\{ \sum_i \varphi_i^{*\alpha}\left( \tau \right) 
\left( \frac{\partial}{\partial \tau} - \mu \right) \varphi_i^{\alpha}\left(\tau \right)  - 
t \sum_{\langle i,j \rangle} \Big( 
\varphi_i^{*\alpha}\left(\tau \right) \varphi_j^{\alpha} \left( \tau \right) + 
\varphi_j^{*\alpha}\left( \tau \right) \varphi_i^{\alpha}\left( \tau \right)
\Big) \right\} \\
\nonumber&&-\frac{v_o^2t^2}{2} 
\sum_{\alpha,\gamma} \sum_{i} 
\int_0^\beta d\tau \int_0^\beta d\tau^\prime 
\varphi^{* \alpha}_i \left( \tau \right) \varphi^{\alpha}_i\left( \tau \right) 
\varphi^{* \gamma}_i \left(\tau^\prime \right) \varphi^{\gamma}_i \left( \tau^\prime \right).
\end{eqnarray}
As one can see from (\ref{effective-action}), in the replicated action the disorder enters as an 
effective attractive interaction between replicas with strength given by $v_o^2t^2/2$.

\section{Shift of the Critical Temperature}

In this Section we determine the shift $\delta T_c \equiv T_c - T_c^{(0)}$ 
of the critical temperature for non-interacting bosons in a three-dimensional lattice 
with random on-site energies. To do this, we use the replicated action (\ref{effective-action}) to 
write down a perturbative expansion in $v_o^2$ of the Green functions of the disordered system: this allows to compute 
both the shift of the chemical potential induced by the disorder and 
the shift $\delta T_c$.

The relevant Green functions are 
\begin{eqnarray}
\label{G1}
&&{\cal G}_{ij}(z)=\langle\varphi_i\varphi_j^*\rangle_{v_o=0}=\int_{BZ}\frac{d\bf k}{(2\pi)^3}\frac{e^{\i{\bf k}\cdot(i-j)}}{E_{\bf k}-z}\\
&&{G}_{ij}(z)=\langle\varphi_i\varphi_j^*\rangle_{v_o\neq 0}
\label{G2}
\end{eqnarray}
where $z=\mu+i\omega_n$ and $\omega_n$ are the bosonic Matsubara frequencies. With ${\cal G}$ ($G$) we denote 
the Green function when there is not (there is) disorder. 
Upon introducing the self-energy function ${\sf \Sigma}({\bf k},z)$, 
one may write the Dyson equation in Fourier space as 
\begin{equation}
G^{-1}({\bf k},z)={\cal G}^{-1}({\bf k},z)+[{\sf \Sigma}({\bf k},z)-\delta\mu],
\end{equation}
where the Fourier transform of ${\cal G}$ is given by 
\begin{equation}
{\cal G}^{-1}({\bf k},z)=E_{\bf k}-z
\end{equation}
and $\delta\mu$ (the shift of the chemical potential) is defined as
\begin{equation}
\delta \mu\equiv {\sf \Sigma}({\bf k}=0,z=\mu).
\end{equation}

Bose-Einstein condensation occurs when 
\begin{equation}
G({\bf k},z)^{-1}\Big{|}_{{\bf k}=0,z=\mu}=-6t-\mu+[{\sf \Sigma}(0,\mu)-\delta\mu]=0
\label{cond-BEC}
\end{equation}
which, at zeroth order in $v_o^2$, is solved by
\begin{displaymath}
\mu_c=-6t:
\end{displaymath}
thus, at any order, the chemical potential at $T_c$ is given by 
$\mu_c+\delta\mu_c=-6t+{\sf \Sigma}(-6t)$. 


To compute Eq.~(\ref{G2}) at the first order in $v_o^2$, 
one needs to evaluate the first terms of the Taylor expansion of $e^{-S_{eff}}$. Using Wick's theorem one gets
\begin{equation}
\label{contractions}
\frac{v_o^2t^2}{2}\sum_{i, \alpha, \gamma}
\langle \varphi_l^1\varphi_m^{*1}\,\varphi^{* \alpha}_i \varphi^{\alpha}_i 
\varphi^{* \gamma}_i \varphi^{\gamma}_i\rangle_{0}=v_o^2t^2
\sum_{i, \alpha, \gamma} \Big\{
\langle \varphi_l^1 \varphi^{* \alpha}_i\rangle _{0}
\langle\varphi^{\alpha}_i  \varphi^{* \gamma}_i \rangle_{0}
\langle\varphi^{\gamma}_i  \varphi^{*1}_m \rangle_{0}+
\langle \varphi_l^1 \varphi^{* \alpha}_i\rangle _{0}
\langle\varphi^{\gamma}_i  \varphi^{* \gamma}_i \rangle_{0}
\langle\varphi^{\alpha}_i  \varphi^{*1}_m \rangle_{0} \Big\}.
\end{equation}
The quantities 
\[\langle \varphi_i^{\alpha} \varphi^{* \gamma}_i\rangle _{0}=\delta_{\alpha\gamma}\,{\cal G}_{ij}\]
are diagonal in the replica indices. 
From Eq.~(\ref{contractions}) one gets
\begin{eqnarray}
G_{lm} = {\cal G}_{lm}+{v_o^2t^2}\,\sum_{i}(1+N)\,
{\cal G}_{li}\,{\cal G}_{ii}\,{\cal G}_{im}+O(v_o^4),
\label{final_N}
\end{eqnarray}
which for $N\rightarrow 0$ leads to
\begin{eqnarray}
G_{lm} = {\cal G}_{lm}+{v_o^2t^2}\,\sum_{i} \,
{\cal G}_{li}\,{\cal G}_{ii}\,{\cal G}_{im}+O(v_o^4).
\label{final}
\end{eqnarray}
From Eq.~(\ref{final}) and using Eq.~(\ref{G1}) one gets 
for the self-energy:
\begin{equation}
\label{self}
{\sf \Sigma}_1({\bf k},z)\equiv {\sf \Sigma}_1(z) =-v_o^2t^2 {\cal G}_{ii}= -v_o^2t^2 \int_{BZ} 
\frac{d{\bf k}^\prime}{(2\pi)^3}\,\frac{1}{E_{{\bf k}^\prime}-z} \equiv
-v_o^2 t^2 \, {\sf F}(z), 
\end{equation}
where the subscript $_1$ denotes that we are including contributions 
up to the order $v_o^2$; furthermore  
\begin{equation}
\label{S1}
{\sf F}(z)=-\frac{1}{z}\,{\sf W}\left( -\frac{z}{6t} \right)
\end{equation}
with ${\sf W}(z)$ defined in Eq.~(\ref{watson}). 
From Eq.~(\ref{self}) one may readily compute at the critical point the shift of the chemical 
potential $\delta \mu_{c1}$ at the first order in $v_o^2$: 
one finds
\begin{equation}
\delta\mu_{c1}=-v_o^2t^2\,{\sf F}(-6t)=v_o^2 t \, \frac{{\sf W}(1)}{6}\approx 0.25\,v_o^2 t.
\end{equation}

As one can see from Eq.~(\ref{cond-BEC}), due to 
the contribution given by the self-energy to the chemical potential, 
the filling fraction $f$ is changed in presence of disorder to $f+\delta f_1$, 
with $\delta f_1$ given by
\begin{eqnarray}
\nonumber\delta f_1 &=& \lim_{\tau\rightarrow 0^-} T_c\sum_n e^{-i\omega_n\tau}
\int_{BZ} \frac{d{\bf k}}{(2\pi)^3} \left\{\frac{1}{E_{\bf k}-z+{\sf \Sigma}_1(z)-\delta\mu_{c1}}- \frac{1}{E_{\bf k}-z}\right\}\Big|_{z=\mu_c+i\omega_n}\\
&\simeq& -\lim_{\tau\rightarrow 0^-}T_c\sum_n e^{-i\omega_n\tau}
\int_{BZ} \frac{d{\bf k}}{(2\pi)^3} \frac{{\sf \Sigma}_1(z)-\delta\mu_{c1}}{(E_{\bf k}-z)^2}\Big|_{z=\mu_c+i\omega_n}.
\label{deltan}
\end{eqnarray}
Substituting 
\begin{equation}
\label{def_dF}
\int_{BZ} \frac{d{\bf k}}{(2\pi)^3}\,\frac{1}{(E_{\bf k}-z)^2}=\frac{\partial}{\partial z} {\sf F}(z)
\end{equation}
in Eq.~(\ref{deltan}), one gets
\begin{equation}
\delta f_1 = \lim_{\tau\rightarrow 0^-}T_c\sum_n e^{-i\omega_n\tau}{v_o^2t^2}\left\{\Big[{\sf F}(z)-{\sf F}(\mu_c)\Big]\frac{\partial}{\partial z} {\sf F}(z)
\right\}\Big{|}_{z=\mu_c+i\omega_n}.
\label{dn_h}
\end{equation}
For large values of the filling $f$, the dominant contribution in Eq.~(\ref{dn_h})
is given by the lowest Matsubara frequency, which yields
\begin{equation}
\delta f_1=v_o^2\, t^2 \, T_c \lim_{z\rightarrow \mu_c}
\left\{\Big[{\sf F}(z)-{\sf F}(\mu_c)\Big]\frac{\partial}{\partial z} {\sf F}(z)\right\}.
\label{dn_h_0}
\end{equation} 

For an explicit computation of (\ref{dn_h_0}), it is most useful to define the dimensionless 
variable $\tilde{z}=z/t$ and the function $\tilde{{\sf F}}(\tilde{z})=-(1/\tilde{z}) 
{\sf W}(-\tilde{z}/6)$. Since $\mu_c/t=-6$, one has 
\begin{equation}
\delta f_1=A \, v_o^2 \frac{T_c}{t} 
\label{dn_h_0_ad}
\end{equation} 
where the numerical coefficient $A$ is given by
\begin{equation}
A=\lim_{\tilde{z}\rightarrow -6^-}
\left\{\Big[\tilde{{\sf F}}(\tilde{z})-
\tilde{{\sf F}}(-6)\Big]\frac{\partial}{\partial \tilde{z}} \tilde{{\sf F}}(\tilde{z})\right\}.
\label{A}
\end{equation} 

In order to compute $A$, one may use the expressions of $\tilde{{\sf F}}$ and ${\sf W}$ in terms 
of elliptic integrals given in \cite{joyce72}: it is 
\begin{equation}
{\sf W}(\tilde{z})=\left( \frac{2}{\pi} \right)^2 \,  
\sqrt{\frac{1-\frac{3}{4}x_1}{(1-x_1)^2}} \, K(k_+) \, K(k_-),
\label{P}
\end{equation} 
where $K(k)$ is the complete integral of the first kind \cite{abramowitz64}, 
$x_1=x_1(\tilde{z})=1/2+1/6\tilde{z}^2-(1/2)\sqrt{(1-1/\tilde{z}^2)
(1-1/9\tilde{z}^2)}$ and 
$k_\pm^2=k_\pm^2(\tilde{z})=1/2 \pm (1/4) 
x_2 \sqrt{4-x_2}-(1/4) (2-x_2) \sqrt{1-x_2}$ with $x_2=x_1/(x_1-1)$. 

One sees that, although 
the function $\partial \tilde{{\sf F}} / \partial \tilde{z}$ 
diverges for $\tilde{z}\rightarrow -6^-$, 
$[\tilde{{\sf F}}(\tilde{z}) - \tilde{{\sf F}}(-6) ] \cdot \partial \tilde{{\sf F}} / \partial \tilde{z}$, yields a finite 
value of $A$. Using (\ref{P}), one may expand the functions 
$\tilde{{\sf F}}$ and $\partial \tilde{{\sf F}} / \partial \tilde{z}$
in terms of $\eta^2\equiv \tilde{z}^2-36\rightarrow 0^+$. One gets
\begin{equation}
{\sf W}(-\tilde{z}/6)
={\sf W}(1)-\frac{3\sqrt{3}}{2\pi}\sqrt{1-\frac{36}{\tilde{z}^2}}
+O(\eta^2),
\end{equation}
\begin{equation}
\tilde{{\sf F}}(\tilde{z})= -\frac{1}{\tilde{z}}
\left({\sf W}(1)-
\frac{3\sqrt{3}}{2\pi}\sqrt{1-\frac{36}{\tilde{z}^2}}\right)+O(\eta^2)
\label{S1-ex}
\end{equation}
and 
\begin{equation}
\label{dS1}
\frac{\partial}{\partial \tilde{z}}\tilde{{\sf F}}(\tilde{z})=
\left(\frac{3\sqrt{3}}{2\pi}\frac{36}{\tilde{z}^4}\frac{1}{\sqrt{1-\frac{36}{\tilde{z}^2}}}
\right)+O(1).
\end{equation}
Inserting Eqs.~(\ref{S1-ex})-(\ref{dS1}) in Eq.~(\ref{A}), 
one readily finds
\begin{equation}
A=-\frac{1}{32\pi^2}.
\label{df}
\end{equation}

As a result of Eq.~(\ref{Tc0}), one notices that, for large $f$, there is a linear relation between $T_c^{(0)}$ and $f$, 
which is given by
\begin{equation}
\frac{{\sf W}(1)} {  6 t} \, T_c^{(0)}=f.
\label{equating-1}
\end{equation}
At the order $v_o^2$, taking into account the 
shift in the particle density given by Eq.~(\ref{df}), one has
\begin{equation}
\frac{{\sf W}(1) } {  6 t}   \, T_c+\delta f_1=f.
\label{equating-2}
\end{equation}
From Eqs.~(\ref{equating-1}) and (\ref{equating-2}) and using Eq.~(\ref{df}), 
one obtains 
\begin{equation}
\label{3D_Tc(v)}
T_c=T_c^{(0)}\left(
1+\frac{3}{16\pi^2{\sf W}(1)}\,v_o^2\right),
\end{equation}
from which
\begin{equation}
\label{3D_dTc(v)}
\frac{\delta T_c}{T_c^{(0)}} = \frac{3 v_o^2}
{16\pi^2{\sf W}(1)} \approx 0.0125 \,v_o^2. 
\end{equation}


To go beyond the first order results, one may 
be tempted to 
use the approach based on the momentum space 
renormalization of the vertex function: 
within ladder approximation, the renormalized vertex 
${\cal V}({\bf k},z_1,z_2)$ can be written for $z_1=z_2\equiv z$ 
as
\begin{equation}
\label{vertex}
{\cal V}({\bf k},z)=\frac{v^2_o t^2}{1-v^2_o t^2\, {\Pi}({\bf k},z)},
\end{equation}
where the polarization function is given by
\begin{equation}
{\Pi}({\bf k},z)=\int_{BZ} \frac{d{\bf q}}{(2\pi)^3}
\frac{1}{(E_{\bf q}-z)(E_{\bf q+k}-z)}.
\end{equation}
As a result, the self-energy ${\sf \Sigma}(z)$ is given by 
\begin{equation}
{\sf \Sigma}(z)=
-\int_{BZ} \frac{d{\bf k}^\prime}{(2\pi)^3}
\frac{{\cal V}(0,z)}{E_{\bf k^\prime}-z},
\end{equation}
and it contributes to the equation for the filling fraction as
\begin{equation}
\label{fil}
f = \lim_{\tau\rightarrow 0^-} T_c\sum_n e^{-i\omega_n\tau}
\int_{BZ} \frac{d{\bf k}}{(2\pi)^3} \left\{\frac{1}{E_{\bf k}-z+{\sf \Sigma}
(z)-\delta\mu_{c}}\right\}\Big|_{z=\mu_c+i\omega_n}.
\end{equation}
At the first order in $v_o^2$, 
the vertex function (\ref{vertex}) is simply given by $v_o^2 t^2$ and the 
corresponding self-energy is 
${\sf \Sigma}(z)=-v_o^2 t^2{\sf F}(z)$, as given by (\ref{self}): 
using Eq.~(\ref{fil}), 
one, of course, retrieves for the shift of the critical 
temperature the result (\ref{3D_dTc(v)}). 
If one attempts to compute the contribution to $f$ 
coming from higher powers in $v_0$, one is immediately faced with problems 
arising from infrared divergences. In fact,
\[
\Pi({\bf k}=0,z)=\frac{\partial {\sf F}(z)}{\partial z}
\]
diverges at $z=\mu_c$. 
The situation is 
similar to what happens for the computation of 
the shift of the critical temperature due to 
weak repulsive interactions, where the perturbation theory 
also fails to 
give a finite expression for $\delta T_c$, 
as discussed in \cite{blaizot08}. 

\section{Conclusions}

In this paper we investigated the effect of random on-site energies on the critical temperature of a non-interacting 
Bose gas on a lattice. By using the replicated action obtained by averaging on the disorder, 
we performed a perturbative expansion for the Green functions of the disordered system. 
We computed, for $v_o^2 \ll 1$, the shift of the chemical potential induced by the 
disorder and the critical temperature for condensation. 
For large filling, the critical temperature is slightly enhanced with respect to the situation in absence of disorder: 
the relative shift $\delta T_c/T_c^{(0)}$ is given by $\delta T_c=a v_o^2$, where the numerical 
coefficient $a$ is positive and small ($a \approx 0.0125$). 

{\em Acknowledgements:} Discussions with 
D. M. Basko and G. Modugno are warmly acknowledged. 
L.D.A., P.S. and A.T. thank the Galileo Galilei Institute for Theoretical 
Physics for the hospitality and INFN for partial support during the completion 
of this work. This work is partly supported by the MIUR project 
``Quantum Noise in Mesoscopic Systems''.

\end{document}